\documentclass[aps,pre,amsmath,amssymb,floatfix,twocolumn,showpacs]{revtex4-1}
\pdfoutput=1
\usepackage{hyperref}
\hypersetup{bookmarksnumbered=true}
\usepackage{graphicx}
\usepackage{bm}
\newcommand{\vect}{\mathbf}

\newcommand{\eqnref}[1]{Eq.~(\ref{#1})}
\newcommand{\figref}[1]{Fig.~\ref{#1}}
\DeclareMathOperator{\sgn}{sgn}
\begin{document}
\begin{abstract}
The linearized potential of a moving test charge in a one-component fully degenerate fermion plasma is studied using the Lindhard dielectric function. 
The motion is found to greatly enhance the Friedel oscillations behind the
charge, especially for velocities larger than a half of the Fermi velocity, in
which case the asymptotic
behavior of their amplitude changes from $1/r^3$ to $1/r^{2.5}$.
In the absence of the quantum recoil (tunneling) the potential reduces to a form similar to that in a classical Maxwellian plasma, 
with a difference being that the plasma oscillations behind the charge at velocities larger than the Fermi velocity are not Landau-damped.
\end{abstract}
\title{Shielding of a moving test charge in a quantum plasma}
\author{D. Else}
\affiliation{School of Physics,
The University of Sydney, New South Wales 2006, Australia}

\author{R. Kompaneets}
\affiliation{School of Physics,
The University of Sydney, New South Wales 2006, Australia}

\author{S. V. Vladimirov}
\affiliation{School of Physics, The University of Sydney,
New South Wales 2006, Australia}

\date{August 31, 2010}
\pacs{52.25.Mq, 52.35.Fp, 03.75.Ss}
\maketitle
\section{Introduction}
The Debye shielding of a moving test charge in a classical plasma is one of the most fundamental 
problems in plasma physics \cite{neufeld-rit-phys.rev-1955, oppenheim-kam-phys.fluids-1964, joyce-mon-phys.fluids-1967,
montgomery-joy-sug-plasma.phys-1968, cooper-phys.fluids-1969, laing-lam-fie-j.plasma.phys-1971, chenevier-dol-per-j.plasma.phys-1973,
stenflo-yu-shu-phys.fluids-1973, wang-joy-nic-j.plasma.phys-1981, peter-j.plasma.phys-1990, trofimovich-kra-sov.phys.jetp-1993,
vladimirov-nam-phys.rev.e-1995, vladimirov-ish-phys.plasmas-1996,ishihara-vla-phys.plasmas-1997,lampe-joy-gan-phys.plasmas-2000}. 
In the collisionless case the motion is known to result in the $1/r^3$-dependence of the potential
at large distances \cite{montgomery-joy-sug-plasma.phys-1968}. The angular
dependence of this asymptotic form is determined by the velocity distributions of the plasma components 
\cite{montgomery-joy-sug-plasma.phys-1968}, and, for instance,
for a one-component Maxwellian plasma the potential is repulsive (for a particle
of like charge) in front of the test charge
and attractive behind it and perpendicular to the motion \cite{cooper-phys.fluids-1969,peter-j.plasma.phys-1990,kompaneets-vla-ivl-new.j.phys-2008}.
In addition, behind the charge a large number of potential minima is formed at substantially suprathermal velocities \cite{neufeld-rit-phys.rev-1955, peter-j.plasma.phys-1990,kompaneets-vla-ivl-new.j.phys-2008}. 
The presence of collisions can lead to the $1/r^2$-dependence of the potential \cite{stenflo-yu-shu-phys.fluids-1973}, 
while the excitation of ion-sound waves can result in the formation of an oscillatory wake structure inside
the corresponding Mach cone \cite{vladimirov-nam-phys.rev.e-1995, vladimirov-ish-phys.plasmas-1996,ishihara-vla-phys.plasmas-1997}.

While the above studies deal with classical plasmas, recently there has been a rapidly growing interest in quantum plasmas, 
motivated primarily by the development of nanostructured metallic and semiconductor materials
\cite{haas-man-fei-phys.rev.e-2000, manfredi, kremp-book,
manfredi-her-appl.phys.lett-2007, melrose-book, ali-mos-kou-new.j.phys-2008, brodin-mar-man-phys.rev.lett-2008, shukla-nature.phys-2009}.
Much attention has been given to a one-component weakly coupled fully degenerate fermion plasma \cite{goldman-zh.eksp.teor.fiz-1947,silin-zh.eksp.teor.fiz-1952,
klimontovich-sil-zh.eksp.teor.fiz-1952,bohm-pin-phys.rev-1953,klimontovich-sil-sov.phys.usp-1960,
physkin,principles,krivitskii-vla-zh.eksp.teor.fiz-1991,vladimirov-phys.scr-1994,manfredi-haa-phys.rev.b-2001,
melrose-mus-phys.plasmas-2009}.
Such a plasma is often described by the Lindhard dielectric function \cite{lindhard,physkin}
where the only quantum effects included are the degeneracy and the quantum recoil (tunneling).
It can be derived using the
Wigner-Poisson system \cite{wigner,tatarskii-sov.phys.usp-1983,hillery-con-scu-phys.rep-1984,diner-transp.theor.stat.phys-1997,
manfredi,haas-man-phys.rev.e-2001,manfredi-haa-phys.rev.b-2001}, a quantum analog of the Vlasov-Poisson system.

We present an investigation of the shielding of a moving classical charge in a quantum plasma
using the Lindhard dielectric function.
The free parameters in this model are 
the velocity of the test charge in units of the Fermi velocity and the plasma coupling parameter.
The latter governs the role of the quantum recoil and should be small
for the model to apply, as discussed in Sec.~\ref{discussion}. 
In the limit of zero coupling parameter (i.e.\ in the absence of the quantum recoil) the potential
is semiclassical in the sense that it can be found using the classical approach but with
a degenerate velocity distribution \cite{manfredi-haa-phys.rev.b-2001}. Such a potential has 
the aforementioned $1/r^3$-asymptote at large distances and was investigated in Ref.~\cite{shukla-ste-bin-phys.lett.a-2006}
at small velocities.
At a finite coupling parameter but zero velocity the potential is known to have Friedel oscillations \cite{physkin,grassme-bus-phys.lett.a-1993}.
Our study extends these results to the general case of arbitrary velocity and coupling parameter and
shows how the quantum recoil changes the semiclassical potential and
how the motion modifies the Friedel oscillations. 

\section{\label{section_model}Model}
The potential around a point test charge $Q$ moving at a constant
velocity $\vect{v}$ in a three-dimensional plasma is given in the linear approximation by the formula \cite{montgomery-joy-sug-plasma.phys-1968,physkin,ivlev-khr-zhd-phys.rev.lett-2004}
\begin{equation}
\label{screening_formula}
\varphi(\vect{r}) = \frac{Q}{4\pi\epsilon_0}\frac{1}{2\pi^2}
\int \frac{\exp(i\vect{k}\cdot\vect{r})}{k^2 D(\vect{k} \cdot \vect{v},
\vect{k})} d\vect{k},
\end{equation}
where $\vect{r}$ denotes the position relative to the instantaneous position
of the charge, $\epsilon_0$ is the electric constant, and $D(\omega, \vect{k})$ is the dielectric
function of the plasma. We take the screening to be due to the response of a single
fully degenerate plasma component (e.g. electrons), with all the other components
remaining fixed as a homogeneous neutralizing background. To describe their
response, we use the Lindhard dielectric function \cite{klimontovich-sil-zh.eksp.teor.fiz-1952, lindhard,klimontovich-sil-sov.phys.usp-1960, physkin, krivitskii-vla-zh.eksp.teor.fiz-1991}:
\begin{equation}
\label{lindhard}
D(\omega, \vect{k}) = 1 + \frac{m\omega_p^2}{\hbar n k^2} \int
\frac{f(\vect{p} + \hbar\vect{k}/2) - f(\vect{p} - \hbar\vect{k}/2)}
{\omega + i\nu - \vect{k} \cdot \vect{p}/m} d\vect{p},
\end{equation}
where $\omega_p=\sqrt{ne^2/(\epsilon_0 m)}$ is the plasma frequency,
$n$ is the particle number density (of the component that responds to the test charge),
$e$ and $m$ is their charge and mass, respectively, $\nu$ is an infinitesimal positive number (i.e.\ the limit $\nu \to 0^+$ should be taken),
$f(\vect{p})$ is the
three-dimensional Fermi-Dirac distribution function:
\begin{equation}
\label{fermidirac_fully_degenerate}
f(\vect{p}) = \begin{cases}
2\left(\dfrac{1}{2\pi\hbar}\right)^3 & \text{if $|\vect{p}| < p_F$}, \\
0 & \text{if $|\vect{p}| > p_F$}, \\
\end{cases}
\end{equation}
$p_F=\hbar (3 \pi^2 n)^{1/3}$ is the Fermi momentum, and $\hbar$ is the Planck constant over $2\pi$.
The applicability of the model is discussed in Sec.~\ref{discussion}.

Carrying out the integral in \eqnref{lindhard} gives \cite{lindhard, physkin}:
\begin{subequations}
\label{dielectric_quantum}
\begin{multline}
D(\omega, \vect{k}) = 1 + \\
\frac{1}{k^2\lambda_{TF}^2}\frac{1}{2a}\left[ F(\omega +
i\nu + a, k) - F(\omega + i\nu - a, k) \right],
\end{multline}
where
\begin{gather}
a = \frac{\hbar k^2}{2m}, \\
F(\Omega, k) = \frac{\Omega}{2} + \frac{(kv_F)^2 - \Omega^2}{4kv_F} 
\ln\left(\frac{\Omega + kv_F}{\Omega - kv_F}\right),
\end{gather}
\end{subequations}
$v_F = p_F/m$ is the Fermi velocity, and $\lambda_{TF} = v_F/(\omega_p\sqrt{3})$ is the Thomas-Fermi screening length; the principal branch of the complex
logarithm should be taken.

Unless otherwise stated, the figures in this paper have been generated by
numerical integration of \eqnref{screening_formula} using the dielectric
function (\ref{dielectric_quantum}). We choose the free parameters to be
the degenerate Mach number $M=v/v_F$ and the parameter
\begin{equation}
\eta = \frac{\hbar \omega_p}{4E_F},
\end{equation}
where $E_F=p_F^2/(2m)$ is the Fermi energy. Note that the parameter $\eta$ is related to the plasma coupling parameter $\Gamma=e^2n^{1/3}/(4\pi \epsilon_0E_F)$
via
\begin{equation}
\eta = \sqrt{\frac{\pi \Gamma}{2(3\pi^2)^{2/3}}}.
\end{equation}
In Appendix \ref{non_dimensional_form}, we have reformulated the problem in non-dimensional form in
terms of the free parameters $\eta$ and $M$.

\section{Results}
\subsection{Semiclassical limit}
\label{semiclassical_limit}
We first consider the semiclassical limit $\eta \to 0$, where the
degeneracy of the unperturbed state is the only quantum effect. The extent to
which the results of the present subsection are relevant despite the fact that we neglect
relativistic effects is discussed
in Sec.\ \ref{discussion}.
In the semiclassical limit we have
\cite{principles}
\begin{equation}
\label{semiclassical_dielectric}
D(\vect{k}\cdot\vect{v}, \vect{k}) = 1 + \frac{1}{k^2 \lambda_{TF}^2}
G(M \vect{\hat{k}} \cdot \vect{\hat{v}}),
\end{equation}
where the hat denotes unit vectors,
\begin{equation}
G(x) = 1 - \frac{x + i\epsilon}{2} \ln\left(\frac{x + i\epsilon + 1}{x + i\epsilon - 1}\right),
\end{equation}
and $\epsilon$ is an infinitesimal positive number.
If we set $v = 0$
we get the exponentially screened potential of the Debye form:
\begin{equation}
\varphi(\vect{r}) = \frac{Q}{4\pi\epsilon_0r}
\exp\left(-\frac{r}{\lambda_{TF}}\right).
\end{equation}

\begin{figure}
\includegraphics{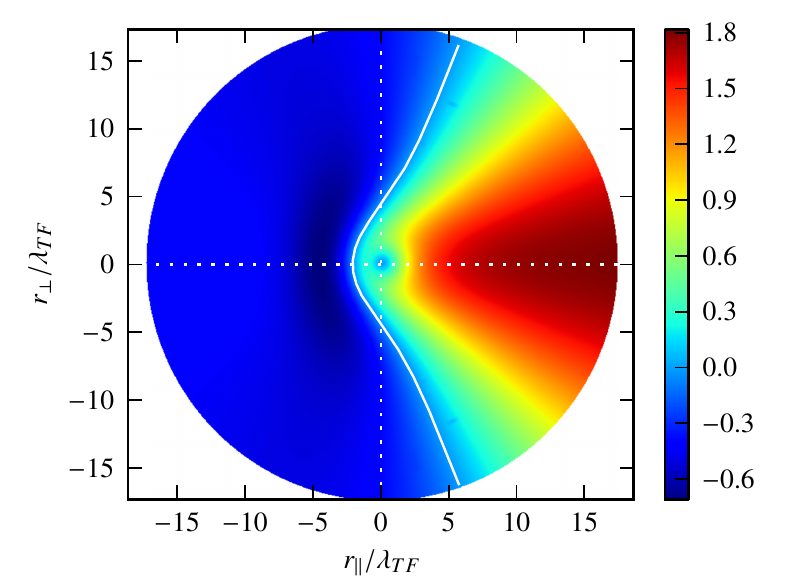}
\caption{\label{semiclassical_polar_v_0_5_cuberatio}The quantity $\varphi r^3$ [in units of $Q\lambda_{TF}^2/(4\pi\epsilon_0)$]
in the semiclassical case ($\eta \to 0$), with the test charge at the origin and
moving at speed $v = 0.5v_F$ to the right. The white superimposed curve denotes
the boundary between positive and negative potential (for $Q > 0$, the positive
potential is on the right of the figure).}
\end{figure}

\begin{figure}
\includegraphics{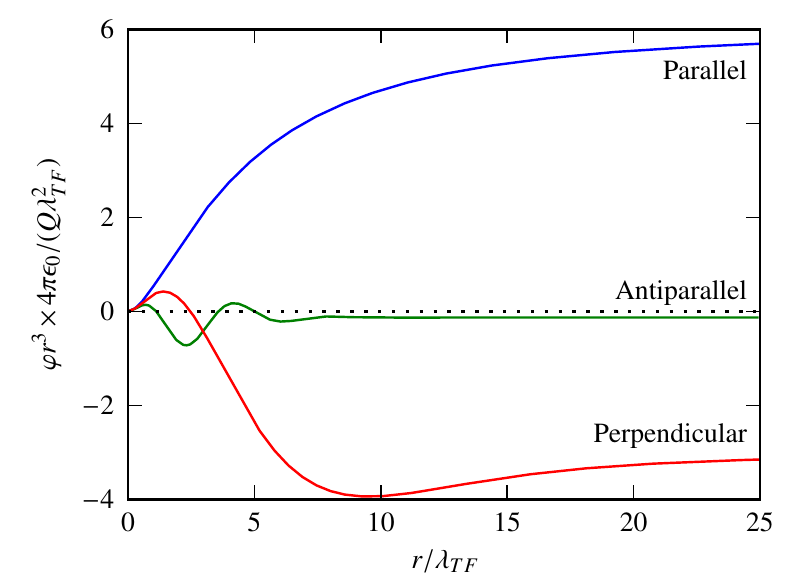}
\caption{\label{semiclassical_v_1_0_three_cuberatio}The quantity $\varphi r^3$ [in units of $Q\lambda_{TF}^2/(4\pi\epsilon_0)$]
in the semiclassical case, with the test charge moving at the Fermi velocity, in
the directions parallel to the motion ($\vect{\hat{r}} \cdot \vect{\hat{v}} =
1$), perpendicular to the motion ($\vect{\hat{r}} \cdot \vect{\hat{v}} = 0$),
and antiparallel to the motion ($\vect{\hat{r}} \cdot \vect{\hat{v}} = -1$).}
\end{figure}

\begin{figure}
\includegraphics{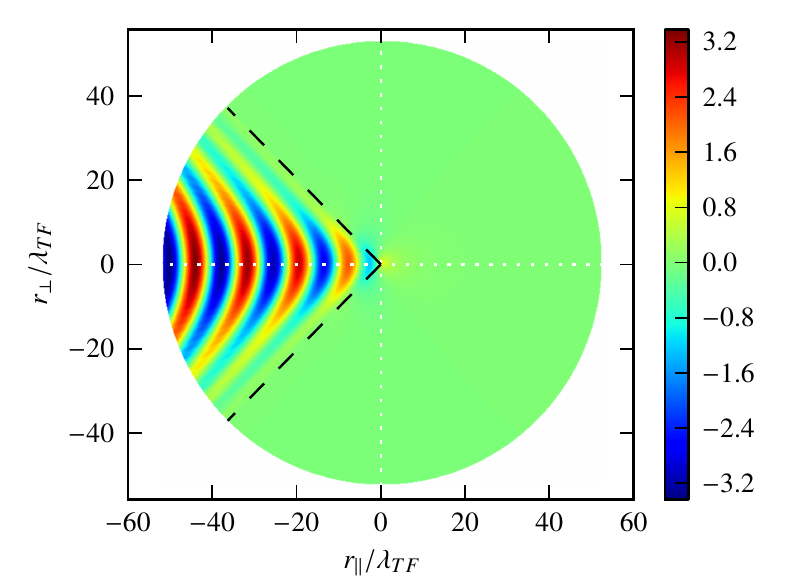}
\caption{\label{semiclassical_polar_v_1_4}The ratio of the potential $\varphi(\vect{r})$ to the unscreened Coulomb potential
$Q/(4\pi\epsilon_0r)$, in the semiclassical case with the test charge at the
origin and moving
at speed $v = 1.4v_F$ to the right. Also shown (dashed lines) is the ``Mach
cone'', defined by $r_{\parallel}/|r_{\perp}| = -\sqrt{M^2-1}$.}
\end{figure}

Now, for $M < 1$, 
the reciprocal dielectric function $1/D(\vect{k} \cdot \vect{v}, \vect{k})$
does not have singularities at real $\vect{k}$
and hence, as shown in
Ref.~\cite{montgomery-joy-sug-plasma.phys-1968}, the asymptotic potential as $r \to \infty$ is
\begin{equation}
\varphi(\vect{r}) = \frac{Q}{4\pi\epsilon_0} H(\gamma)
\frac{\lambda_{TF}^2}{r^3} + O\left(\frac{1}{r^5}\right)
\end{equation}
where $\gamma$ is the angle between $\vect{r}$ and $\vect{v}$, and
\cite{kompaneets-mor-ivl-phys.plasmas-2009}
\begin{multline}
\label{H_of_gamma}
H(\gamma) =  -\frac{i}{\pi^2} \lim_{\delta \to 0^{+}}\int_0^{2\pi} d\phi \int_{-1}^1
d\mu \\
\times \frac{1}{G[M(\mu \cos \gamma + \sqrt{1 - \mu^2} \sin \phi \sin
\gamma)]} \frac{1}{(\mu + i \delta)^3}.
\end{multline}
As $M \to 0$, we can expand $H(\gamma)$ in powers of $M$, giving the asymptotic
formula
\begin{multline}
\label{semiclassical_asymptotic}
\varphi(\vect{r}) =
\frac{Q}{4\pi\epsilon_0r}\exp\left(-\frac{r}{\lambda_{TF}}\right)\\
+ \frac{Q}{4\pi\epsilon_0}\frac{\lambda_{TF}^2}{r^3}
\left[2M \cos \gamma + \left(\frac{\pi^2}{4} - 1\right) M^2 (3\cos^2 \gamma - 1)
\right] \\
+ O\left(\frac{M^3}{r^3}\right) + O\left(\frac{M}{r^5}\right)
\end{multline}
as $r \to \infty$, $M \to 0$. The term linear in $M$ in
\eqnref{semiclassical_asymptotic} was given in \cite{shukla-ste-bin-phys.lett.a-2006}, although they
appear to be missing the factor of 2. This asymptotic result is qualitatively
identical (the only difference is in the numerical coefficients in front of the
two terms) to the classical case of a Maxwellian plasma \cite{cooper-phys.fluids-1969}.
Equation~(\ref{semiclassical_asymptotic}) shows that for small nonzero $M$, an
attractive (for like charges) potential forms antiparallel and perpendicular to the motion, whereas
the potential parallel to the motion remains repulsive (but decays as $1/r^3$
instead of exponentially as for $M = 0$). These features persist at velocities up to and
including the Fermi velocity, as illustrated in Figs.\
\ref{semiclassical_polar_v_0_5_cuberatio} and
\ref{semiclassical_v_1_0_three_cuberatio}.
(Note that although Fig. 1 exhibits a cone-shaped feature in front of the charge, this is not an oscillatory Mach cone 
of the kind that occurs behind fast-moving charges and is shown in Fig.~\ref{semiclassical_polar_v_1_4}.)


On the other hand, for $M > 1$, the reciprocal dielectric function
$1/D(\vect{k}\cdot\vect{v}, \vect{k})$ has a pole for real $\vect{k}$
[the parameter $\epsilon$ sets the rule for avoiding this pole in the integral (\ref{screening_formula})].
Physically, this corresponds to the fact that a test charge moving faster than
the Fermi velocity can excite plasma oscillations. Furthermore, since these
oscillations have phase velocity greater than the Fermi velocity, they are not
Landau-damped, as can be seen mathematically in the fact that $D(\omega,k)$ has
no imaginary part at $\omega/k > v_F$. This is because, with the unperturbed
velocity distribution given by \eqnref{fermidirac_fully_degenerate}, there are
no particles with velocities greater than $v_F$ and hence no particles satisfying the resonance
condition to contribute to Landau damping (this resonance condition is, since the quantum recoil
disappears in the semiclassical limit, the same as in classical
plasma physics) \cite{principles, krivitskii-vla-zh.eksp.teor.fiz-1991}.
The result is a strong oscillatory wake
(with amplitude decreasing as $1/r$) behind the test charge with an infinite number of minima, as shown in
\figref{semiclassical_polar_v_1_4}. Note that  an oscillatory structure behind the charge
is also present in the case of a Maxwellian distribution at substantially suprathermal velocities,
but it is Landau-damped \cite{peter-j.plasma.phys-1990}. As a result, the number
of minima is finite (though very large)
because the $1/r^3$-asymptote falls off slower than the exponentially damped amplitude of oscillations.

\subsection{General case}
\subsubsection{The case of \texorpdfstring{$0 \leq v \leq v_F$}{0 <= v < v\_F}}
\label{general_v_less_than_v_F}
As is well known \cite{physkin}, at nonzero $\eta$
the static dielectric function $D(0, \vect{k})$ has
a non-analyticity (the ``Kohn anomaly'' \cite{kohn}) at wavenumbers $|\vect{k}| = 2k_F$,
where $k_F = p_F/\hbar$ is the Fermi wavenumber. The Kohn anomaly is related
to the discontinuous Fermi surface and gives
rise to the Friedel oscillations \cite{friedel}, with the potential as $r \to
\infty$ given by \cite{physkin}
\begin{equation}
\label{static_friedel}
\varphi(\vect{r}) = \frac{Q\lambda_{TF}^2}{4\pi\epsilon_0} \frac{36\eta^4}{(2 +
3\eta^2)^2} \frac{\cos(2k_F r)}{r^3} + o\left(\frac{1}{r^3}\right).
\end{equation}

\begin{figure}
\includegraphics{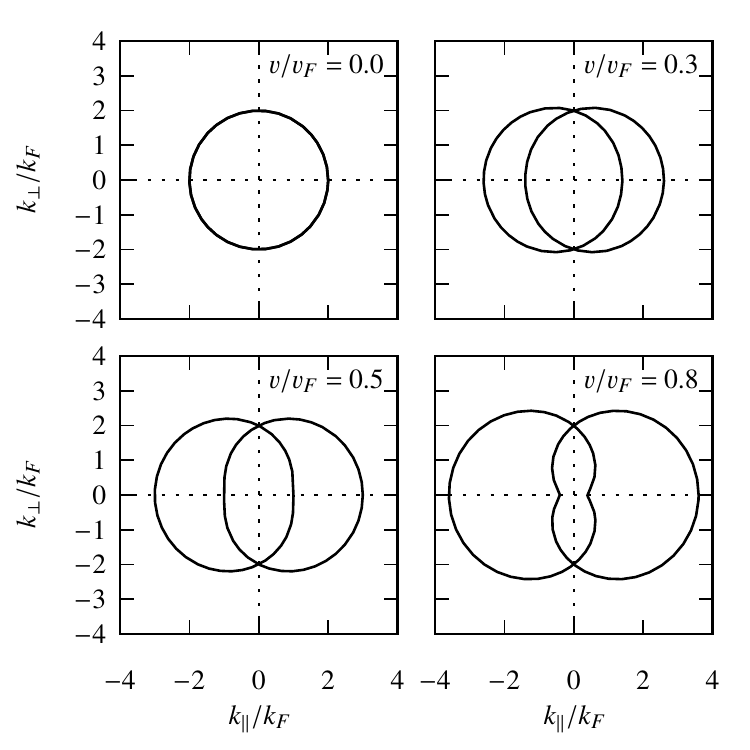}
\caption{\label{kohn_singularity}The wavenumbers $\vect{k}$ at which the Kohn
anomaly in the dielectric function $D(\vect{k}\cdot\vect{v}, \vect{k})$
occurs; the velocity $\vect{v}$ is directed horizontally. This figure is valid for any value of $\eta$ since the axes are
normalized to the Fermi wavenumber $k_F$.}
\end{figure}

For nonzero velocities, the Kohn anomaly occurs at wavenumbers $\vect{k}$
such that
\begin{equation}
\label{kohn_equation}
|\vect{k}| = 2k_F(1 \pm M \vect{\hat{v}}\cdot\vect{\hat{k}}),
\end{equation}
as illustrated in \figref{kohn_singularity}.
The asymptotic potential as $r \to \infty$ is the superposition of the contribution from the Kohn
anomaly and from the small wavenumbers ($\vect{k} \to \vect{0}$); the latter is
identical with the total semiclassical asymptotic potential, since
the semiclassical and general forms of the dielectric function coincide in
the limit $\vect{k} \to \vect{0}$.

\begin{figure}
\includegraphics{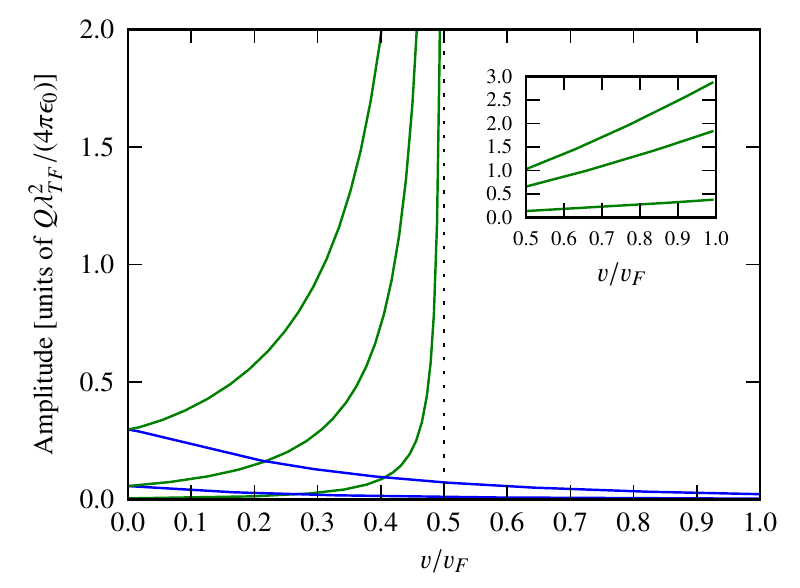}
\caption{\label{plot_amplitudes}
The amplitude of oscillations of the quantity $\varphi r^3$ [in
units of $Q\lambda_{TF}^{2}/(4\pi\epsilon_0)]$, in the limit $r \to \infty$,
behind the charge (green,
increasing), and in front of the charge (blue, decreasing), as a function of
test charge velocity, for values of $\eta$ (in order of increasing amplitude):
0.15, 0.3, and 0.5. (The amplitude for
$\eta = 0.15$ in front of the charge is too small to be visible on this graph). Inset: The
amplitude of oscillations of the quantity $\varphi r^{2.5}$ [in units of $Q
\lambda_{TF}^{1.5}/(4\pi\epsilon_0)$] behind the charge, for $v/v_F > 0.5$.}
\end{figure}

\begin{figure*}
\includegraphics{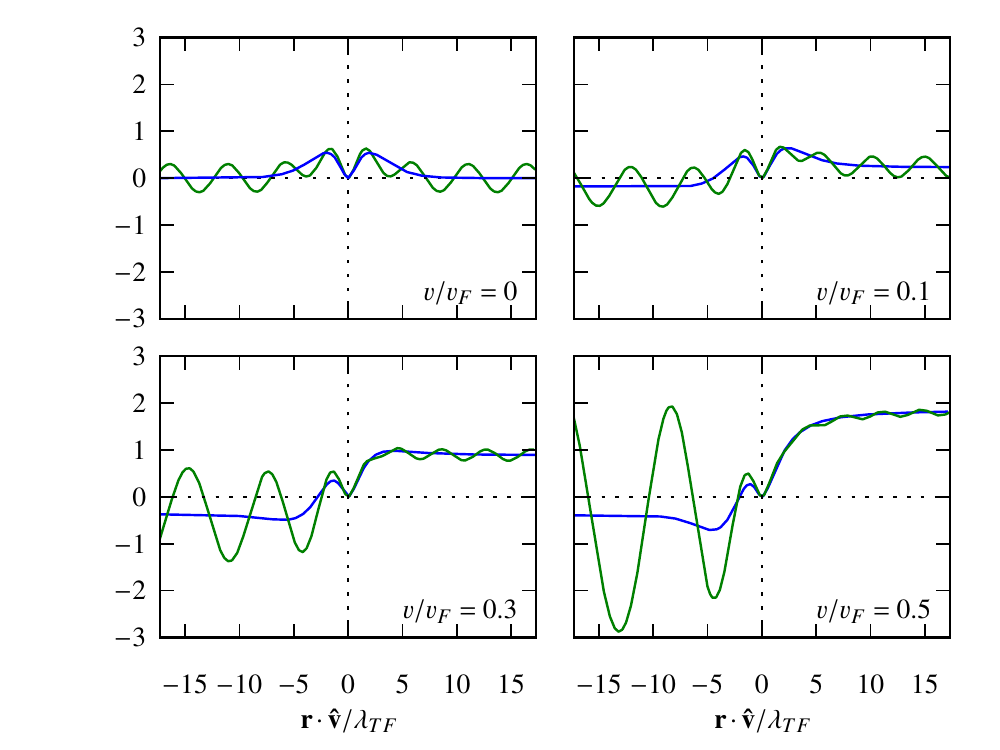}
\caption{\label{cuberatio}The quantity $\varphi r^3$ [in units of
$Q\lambda_{TF}^2/(4\pi\epsilon_0)$], along the direction of motion, in the
semiclassical case (blue, non-oscillatory), and the case $\eta = 0.5$ (green,
oscillatory), at various
velocities of the test charge.}
\end{figure*}

\begin{figure}
\includegraphics{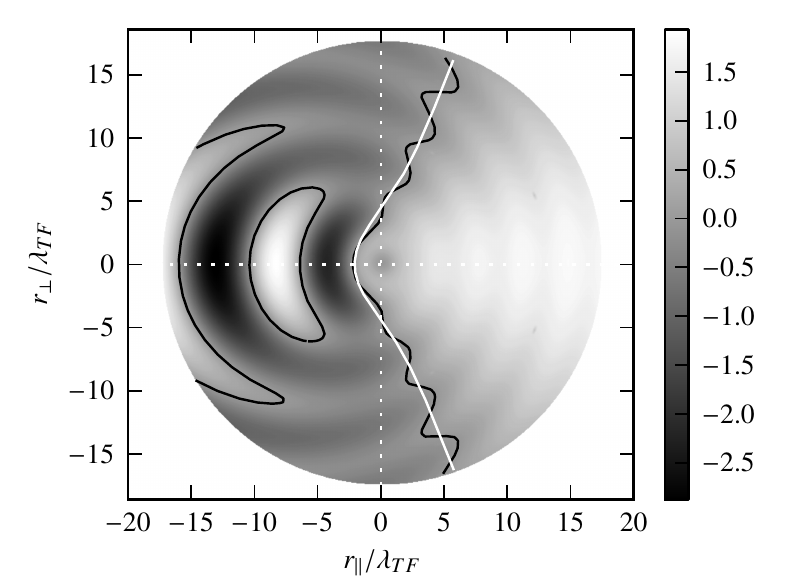}
\caption{\label{quantum_polar_v_0_5}
The quantity $\varphi r^3$ [in units of $Q\lambda_{TF}^2/(4\pi\epsilon_0)$] as a
function of position for $\eta = 0.5$, with the test charge at the origin and moving at speed $v
= 0.5v_F$ to the right. The black and white lines denote the boundary
between positive and negative potential for $\eta = 0.5$ and the semiclassical
case respectively.}
\end{figure}

In Appendix \ref{kohn_asymptotic_derivation}, we outline the derivation of
expressions for the asymptotic contribution resulting from the Kohn anomaly. The
result is, for $\vect{r}$ and $\vect{v}$ parallel:
\begin{multline}
\label{asymptotic_quantum_ahead}
\varphi(\vect{r}) = \frac{Q}{4\pi\epsilon_0 r^3}\biggl\{H(0)\lambda_{TF}^2
\\ + \frac{1}{2k_F \lambda_{TF}^2}
\mathrm{Re}\left[f_{+}(1)\frac{\exp\left[2k_F(1 + M)ir\right]}{1 + 2M} - f_{+}(0)\right]
\biggr\}
\\ + o\left(\frac{1}{r^3}\right),
\end{multline}
where the first term inside the curly brackets represents the semiclassical
asymptotic contribution as described in Sec.\ \ref{semiclassical_limit}, with
$H(\gamma)$ given by \eqnref{H_of_gamma}; and the second term represents the
contribution due to the Kohn anomaly, with
\begin{subequations}
\begin{equation}
f_{\pm}(\mu) = \frac{1}{k_{\pm}^3 \left[D(k_{\pm} v \mu, k_{\pm})\right]^2}
\left[ \mu \mp \frac{2k_F M}{k_{\pm}}(1-\mu^2)\right]^2,
\end{equation}
\begin{equation}
k_{\pm} = 2k_F(1 \pm M|\mu|).
\end{equation}
\end{subequations}
For $\vect{r}$ and $\vect{v}$ antiparallel, we get a similar expression as long
as $M < 1/2$:
\begin{multline}
\varphi(\vect{r}) = \frac{Q}{4\pi\epsilon_0 r^3}\biggl\{H(\pi)\lambda_{TF}^2
\\ + \frac{1}{2k_F \lambda_{TF}^2}
\mathrm{Re}\left[f_{-}(1)\frac{\exp\left[-2k_F(1 - M)ir\right]}{1 - 2M} - f_{-}(0)\right]
\biggr\}
\\ + o\left(\frac{1}{r^3}\right).
\end{multline}
The difference between the two directions arises because, for $\vect{r}$ and
$\vect{v}$ parallel, only the \emph{outer} surface of the Kohn anomaly, i.e.\ 
$|\vect{k}| = 2k_F(1 + M|\vect{\hat{k}}\cdot\vect{\hat{v}}|)$, contributes to the
asymptotic form; whereas, for $\vect{r}$ and $\vect{v}$ antiparallel, only the
\emph{inner} surface, i.e. $|\vect{k}| = 2k_F(1 -
M|\vect{\hat{k}}\cdot\vect{\hat{v}}|)$ contributes.
Finally, for $\vect{r}$ and $\vect{v}$ antiparallel and $1/2 \leq M < 1$, a different asymptotic form emerges, with the
amplitude of oscillations now decaying as $1/r^{2.5}$ rather than $1/r^3$:
\begin{multline}
\label{asymptotic_quantum_behind_stronger}
\varphi(\vect{r}) = \frac{Q}{4\pi\epsilon_0} \sqrt{\frac{\pi}{M}} \frac{1}{2\lambda_{TF}^2\sqrt{k_F}
r^{2.5}} 
\\ \times \mathrm{Re} \left[ (1-i) 
f_{-}\left(\frac{1}{2M}\right) \exp\left(-i\frac{k_Fr}{2M}\right)\right]
\\ + o\left(\frac{1}{r^{2.5}}\right).
\end{multline}
This is related to the change in concavity of the inner surface of the Kohn
anomaly for $M > 1/2$, which can be observed in the last pane of
\figref{kohn_singularity}. The velocity dependence of the amplitude of oscillations derived from Eqs.\
(\ref{asymptotic_quantum_ahead}--\ref{asymptotic_quantum_behind_stronger}) is
shown in \figref{plot_amplitudes}. It can be seen that, with increasing
velocity, the Friedel oscillations become stronger behind the charge (especially
for $M > 1/2$), and weaker in front.
This is qualitatively
illustrated in Figs.\ \ref{cuberatio} and \ref{quantum_polar_v_0_5}. 
Note, however, that no 
substantial change occurs in the Friedel
oscillations in the direction perpendicular to the motion for any $v < v_F$.

\begin{figure}
\includegraphics{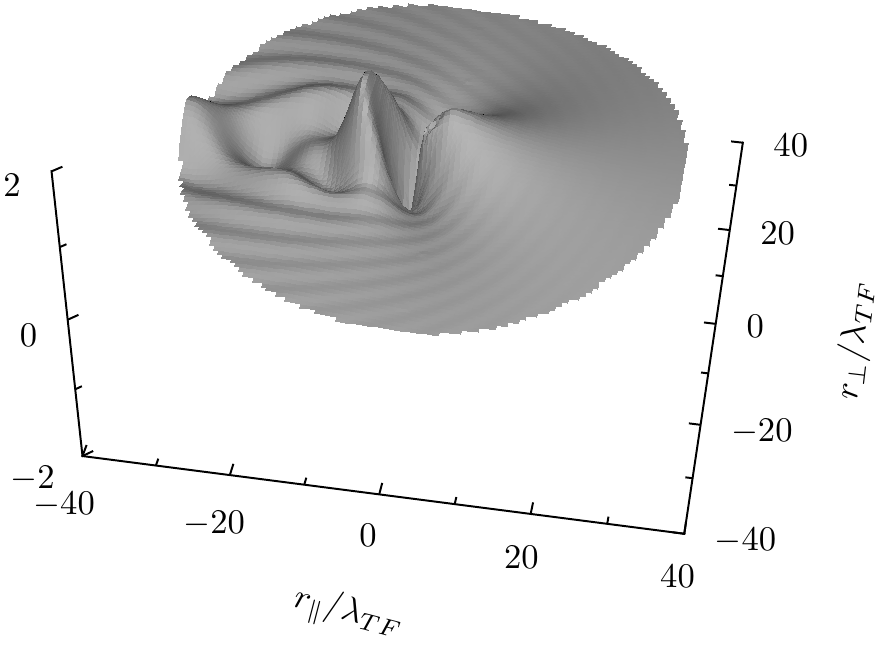}
\caption{\label{v_1_0_surface}
The quantity $\varphi r^2$ [in units of $Q\lambda_{TF}/(4\pi\epsilon_0)$] as a
function of position in the case $\eta = 0.5$, with the test charge is located at the
origin and moving to the right at the Fermi velocity $v_F$.}
\end{figure}

\begin{figure}
\includegraphics{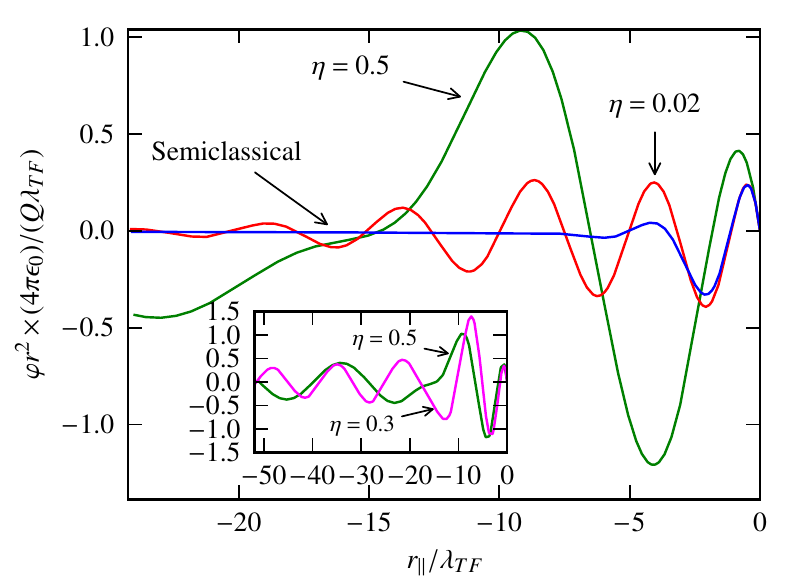}
\caption{\label{v_1_0_comparison}
The quantity $\varphi r^2$ [in units of $Q\lambda_{TF}/(4\pi\epsilon_0)$] behind
the test charge for $v = v_F$. Green is $\eta = 0.5$; red is $\eta = 0.02$; blue
is semiclassical. The test charge is at the right of the figure ($r_{\parallel}
= 0$).
Inset: an extension of the main graph, for $\eta = 0.5$ (green), and $\eta =
0.3$ (magenta).}
\end{figure}

When the test charge is moving at the Fermi velocity (see
Figs.\ \ref{v_1_0_surface} and \ref{v_1_0_comparison}) the stronger
Friedel oscillations are still present behind the test
charge in the far-field, but close to the test
charge a strong irregular wake field of a somewhat different character
emerges. In particular, this near-field wake is still distinctly present even
for very small values of $\eta$ (e.g.\ $\eta = 0.02$).

\subsubsection{The case of \texorpdfstring{$v > v_F$}{v > v\_F}}
\begin{figure}
\includegraphics{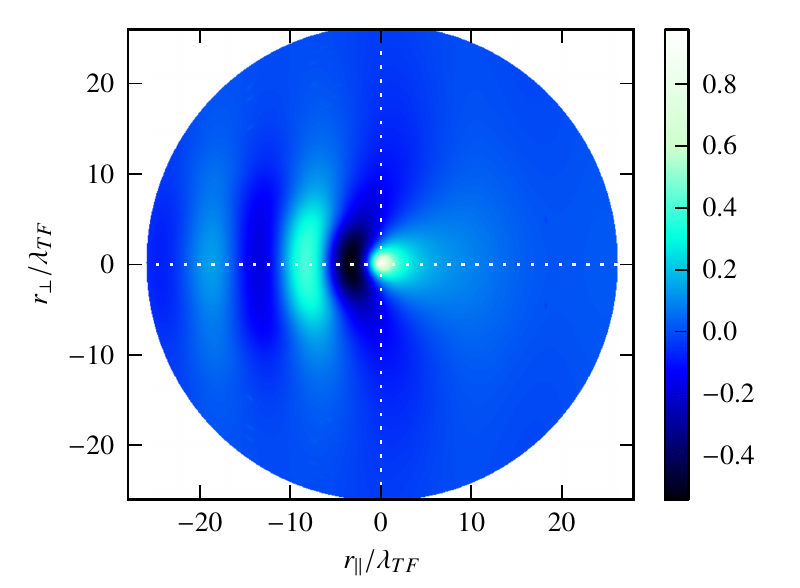}
\caption{\label{quantum_polar_v_1_4}The ratio of the potential $\varphi(\vect{r})$ to the unscreened Coulomb potential
$Q/(4\pi\epsilon_0r)$, in the case $\eta = 0.5$, with the test charge at the
origin and moving
at speed $v = 1.4v_F$ to the right.}
\end{figure}

\begin{figure}
\includegraphics{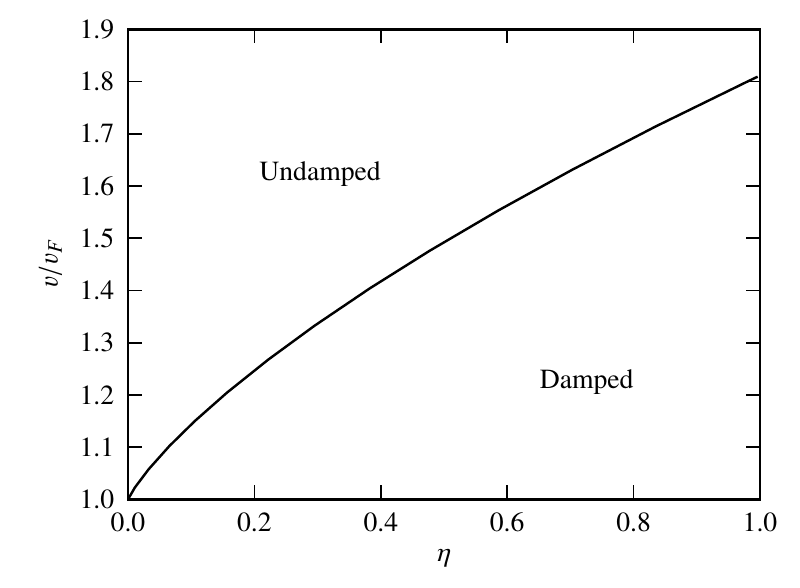}
\caption{\label{plot_thresholds}The dependence of the threshold velocity
$v_{\mathrm{thresh}}$ on the parameter $\eta$. In the region marked ``undamped'',
undamped oscillations are excited. In the region marked ``damped'', they are not
excited. }
\end{figure}


The inclusion of quantum recoil increases the lower bound on the phase velocity
$\omega/k$ of undamped longitudinal oscillations
\cite{krivitskii-vla-zh.eksp.teor.fiz-1991}. This can be shown by numerically
solving the equation $D(\omega,k) = 0$ for real $\omega$ and $k$, where $D(\omega,k)$ is
given by \eqnref{lindhard}, or its equivalent \eqnref{dielectric_quantum}.
The minimum phase
velocity $v_{\mathrm{thresh}}$ is greater than the Fermi velocity, coinciding 
with it in the semiclassical limit, as shown in \figref{plot_thresholds}.
Hence, for test charge velocities
in the range $v_F < v < v_{\mathrm{thresh}}$, the charge cannot excite undamped
oscillations, and the wake depicted in \figref{quantum_polar_v_1_4} is formed. This
is not as strong as in the semiclassical case (it falls off faster than $1/r$),
and is also of a markedly different character, bearing no resemblance to a Mach
cone.

For $v > v_{\mathrm{thresh}}$, on the
other hand, all undamped longitudinal oscillations with phase velocity
$v_{\mathrm{thresh}} < v_{\phi} < v$ are
excited.
We have not performed calculations
for this case, but the results of Ref.~\cite{mazarro-ech-rit-phys.rev.b-1983}
show a strong oscillatory wake, though its asymptotic behavior is
not apparent. For very large
velocities ($v \gg v_F$), the wake
formed will be identical to the semiclassical wake \cite{jakubassa-1977}.

\section{Discussion}
\label{discussion}
Let us discuss the applicability of the model. Firstly, our model is based on the mean-field approximation
which is justified at $\eta \ll 1$ \cite{klimontovich-sil-sov.phys.usp-1960, manfredi,manfredi-haa-phys.rev.b-2001}. 
In the limit $\eta \to 0$, however, the quantum recoil disappears \cite{manfredi-haa-phys.rev.b-2001}. Thus the fact that our model
includes the quantum recoil but does not include particle correlations makes it inconsistent in a certain sense. Nevertheless, it is widely used in the literature,
i.e.\ at small $\eta$ the quantum recoil is assumed to be more important than particle correlations. 
Their effects are discussed, e.g.,
in Refs.~\cite{ferrel-phys.rev-1957, klimontovich-sil-sov.phys.usp-1960,glick-phys.rev-1963,toigo-woo-phys.rev.b-1970,toigo-woo-phys.rev.b-1971}.

Secondly, our model does not include relativistic effects. This imposes a lower bound on the parameter $\eta$ since the latter can be represented as
\begin{equation}
\label{eta_relativistic_condition}
\eta = \sqrt{\frac{\alpha}{3\pi}\frac{c}{v_F}},
\end{equation}
where $c$ is the speed of light and $\alpha=e^2/(4\pi \epsilon_0 \hbar c) \approx 1/137$ is the fine structure constant (here we assumed
that the charge of the particles is equal in its absolute value to the elementary charge).
The condition $v_F \ll c$ requires $\eta \gg \sqrt{\alpha/(3\pi)} \approx
0.03$. Hence the limit of $\eta \to 0$, as considered in Sec.\
\ref{semiclassical_limit}, cannot be taken strictly. Nevertheless, the results
of Sec.\ 
\ref{semiclassical_limit} are relevant in a certain parameter regime. 
Indeed, the limit $\eta \to 0$ corresponds physically to the neglect of the quantum recoil,
while the quantum recoil provides a contribution given approximately by
\eqnref{static_friedel} (for $M \ll 1$ and $r \gg \lambda_{TF}$). This is negligible compared to the
$1/r^3$-term in Eq.~(\ref{semiclassical_asymptotic}) for $5 \eta^4 \ll M$.
(To be exact, the latter inequality is for $|\cos \gamma| \sim 1$, while for $\cos \gamma =0$
it changes to $2 \eta^2 \ll M$.) Thus the applicability range of Eq.~(\ref{semiclassical_asymptotic})
due to the above restriction $\eta \gg 0.03$
is $4 \times 10^{-6} \ll 5 \eta^4 \ll M \ll 1$
(again, it is for $|\cos \gamma| \sim 1$, while for $\cos \gamma =0$
it changes to $2\times 10^{-3} \ll 2 \eta^2 \ll M \ll 1$).
As the velocity increases, the effect of the
quantum recoil becomes less important in front of the charge (such that for
$v/v_F$ not too close to zero the semiclassical and general forms of the
potential are very similar even for $\eta = 0.5$, as can be seen in
\figref{cuberatio}), and more important behind the charge, especially at $v/v_F
> 0.5$.

Thirdly, our model deals with the linearized potential. The linearization is justified when the potential energy of the interaction
of a plasma particle with the test charge at the characteristic screening length (i.e.\ the length at which the Coulomb potential
and the actual potential start to significantly deviate from each other) is much smaller than the characteristic kinetic energy of 
the particles in the frame of the test charge. At small velocities ($v \lesssim v_F$) this condition reads
$|Q e| / (4\pi \epsilon_0 \lambda_{TF}) \ll E_F$ which is equivalent to $12 \pi
(|Q/e|) \eta^3 \ll 1$, while at large velocities ($v \gtrsim v_F$)
the characteristic screening length becomes $v/\omega_p$ and the above condition changes to
$|Q e| \omega_p / (4\pi \epsilon_0 v) \ll m v^2$ which can be rewritten as
$12 \pi (|Q/e|) \eta^3 (v_F/v)^3 \ll 1$. Thus larger velocities imply better applicability of the linear theory.

Finally, there is a number of other effects which are not considered in our model but may be important under certain conditions.
For instance, in our model the shielding is due to one plasma component (e.g.\ electrons) only. The screening in the presence of both electron and ion response
is considered in Refs.~\cite{ali-shu-phys.plasmas-2006, shukla-eli-phys.lett.a-2008} using the quantum hydrodynamic model \cite{manfredi-haa-phys.rev.b-2001}.
Furthermore, we did not include plasma production and loss processes. In classical plasmas they  
can introduce an additional non-screened term and/or a weakly screened
term in the potential and thus change its long-range behavior, even in the absence of the charge motion \cite{fiippov-zag-mom-sov.phys.jetp-2007, 
chaudhuri-khr-mor-phys.plasmas-2008,khrapak-ivl-mor-phys.plasmas-2010, chaudhuri-khr-mor-phys.plasmas-2010}. 
In quantum plasmas these processes (e.g. pair creation and annihilation) can have an important effect on the wave dispersion \cite{mcorist-mel-wei-j.plasma.phys-2007}
and thus can affect the shielding of a moving charge.

The potential of a moving test charge under the Lindhard dielectric function has
previously been investigated in Ref.~\cite{mazarro-ech-rit-phys.rev.b-1983}.
That paper focused on the potential along the line of motion and surrounding the first potential minimum
behind the test charge, for values of
$\eta$ approximately in the range 0.3 -- 0.6, while the present paper is focused on the three-dimensional distribution and asymptotic behavior.
Since the authors of Ref.~\cite{mazarro-ech-rit-phys.rev.b-1983} considered only the ``wake potential'', i.e.\ with the
Coulomb potential subtracted, the actual behavior of the potential 
is not apparent from their work except when the wake is strong enough that
the Coulomb potential is small by comparison, which occurs only when undamped
oscillations are excited, i.e.\ $v >
v_{\mathrm{thresh}}$. We have not studied this case in the present paper.

Our results can be applied, for instance, to investigation
of bound electron states in the wake fields of ions in solids. 
This problem was studied in Ref.~\cite{mazarro-ech-rit-phys.rev.b-1983} using a parabolic fit to the potential around the first potential minimum,
while a detailed investigation requires an accurate knowledge of the whole potential distribution.
Furthermore, our model can be extended to calculate the stopping power or the drag force, similar to classical 
plasmas \cite{peter-mey-phys.rev.a-1991, ivlev-khr-zhd-phys.rev.lett-2004, ivlev-khr-zhd-phys.rev.e-2005,khrapak-ivl-zhd-phys.plasmas-2005,khrapak-cha-mor-ieee.trans.plasma.sci-2009}.

\section{Conclusion}
We have studied the shielding of a moving classical test charge in a fully degenerate fermion plasma 
in both the
semiclassical case (when the degeneracy of the unperturbed state is the only important quantum
effect), and the case when quantum recoil is included.

In the semiclassical case for $v \leq v_F$, the potential goes asymptotically as
$1/r^3$, and is repulsive in front of the test charge, and attractive behind the
test charge and perpendicular to the motion; for $v$ close to $v_F$ the
attraction is especially pronounced perpendicular to the motion. For $v > v_F$
the test charge excites plasma oscillations and a strong oscillatory wake is
formed inside the Mach cone.

We have also found that the inclusion of quantum recoil leads to new effects
entirely absent from the semiclassical case and the case of a Maxwellian distribution.  The Friedel
oscillations, already present in the screening of a static charge, are increased
in strength behind a moving charge, and for $v > v_F/2$
the asymptotic behavior of their amplitude behind the charge changes from
$1/r^3$ to $1/r^{2.5}$. Furthermore, the inclusion of quantum recoil makes the
threshold velocity for excitation of an oscillatory wake of undamped plasma
oscillations larger than the Fermi velocity.

These findings extend the previous results on the shielding of a moving charge in a classical plasma 
to the quantum case and can be applied, for instance, to investigation
of bound electron states in the wake fields of ions in solids.

\begin{acknowledgments}
The authors thank the anonymous referees for helpful comments on the manuscript.
R.~K. acknowledges
the receipt of a Professor Harry Messel Research Fellowship
supported by the 
Science
Foundation for Physics within the University of Sydney.
The work was partially supported by the Australian Research Council. 
\end{acknowledgments}

\appendix
\section{Non-dimensional form of the equations}
\label{non_dimensional_form}
In order to demonstrate that our choice of free parameters $M$ and $\eta$
determines the potential up to
scaling, we express the problem in non-dimensional form in terms of
these parameters. Defining the normalized wavenumber and frequency by 
$\vect{q} \equiv (v_F/\omega_p)\vect{k}$, $w \equiv \omega/\omega_p$,
the dielectric function of \eqnref{dielectric_quantum} can be written as
\begin{subequations}
\begin{multline}
D(\omega,k) = \widetilde{D}(w, q) = 1 + \frac{3}{q^2}\frac{1}{2\eta q^2} \\
\times \left[\widetilde{F}\left(w + i\tilde{\nu} + \eta q^2, q\right)
  - \widetilde{F}\left(w + i\tilde{\nu} - \eta q^2,
  q\right)\right],
\end{multline}
where $\tilde{\nu} = \nu/\omega_p$ is an infinitesmal positive number, and 
\begin{equation}
\widetilde{F}(W,q) = \frac{W}{2} + \frac{q^2 - W^2}{4q} \ln\left(\frac{W + q}{W -
q}\right).
\end{equation}
\end{subequations}
If we now define the normalized variables $\vect{x} \equiv \vect{r}/\lambda_{TF}$,
$\Phi \equiv (4\pi\epsilon_0 \lambda_{TF}/Q)\varphi$, the non-dimensionalized form of
\eqnref{screening_formula} becomes:
\begin{equation}
\Phi(\vect{x}) = \frac{1}{2\pi^2\sqrt{3}}
\int \frac{\exp\left(i \vect{q} \cdot \vect{x}/\sqrt{3}\right)}{q^2 \widetilde{D}(M \vect{q}\cdot\vect{\hat{v}}, \vect{q})}
d\vect{q},
\end{equation}
where $\hat{\vect{v}}$ is the unit vector pointing in the direction of motion of
the test charge.

\section{Derivation of Eqs.\
(\ref{asymptotic_quantum_ahead}--\ref{asymptotic_quantum_behind_stronger})}
\label{kohn_asymptotic_derivation}
In order to derive the asymptotic contribution to the potential from the Kohn
anomaly, we first integrate \eqnref{screening_formula} by parts twice. The
surface terms at $\vect{k} = \vect{0}$ and at infinity vanish; in order to
avoid having to evaluate surface terms at the locations of the Kohn anomaly, we
take the limit $\nu \to 0^{+}$ only after integrating by parts. This leaves:
\begin{multline}
\label{potential_integratedbyparts}
\varphi(\vect{r}) = -\frac{Q}{4\pi\epsilon_0}\frac{1}{2\pi^2 r^2} \\
\times \int \left[\left(\hat{\vect{r}} \cdot \frac{\partial}{\partial
\vect{k}}\right)^2 \left(\frac{1}{k^2 D(\vect{k}\cdot\vect{v},
\vect{k})}\right)\right] \exp(i\vect{k}\cdot\vect{r}) d\vect{k}.
\end{multline}

Assuming now that $\vect{r}$ is parallel or antiparallel to the motion, we
can write 
\begin{equation}
\int d\vect{k} \to 2\pi \int_{-1}^1 d\mu \int_0^\infty k^2 dk,
\end{equation}
where
$\mu = \vect{\hat{k}}\cdot{\vect{\hat{v}}}$. 
Replacing the integral over $k$ by the asymptotic contribution thereto from the
locations of the Kohn anomaly, $k = 2k_F(1 \pm M\mu)$, we obtain
an intermediate expression, which we
denote $\varphi_{\mathrm{kohn}}$, for the asymptotic contribution to the
potential from the
Kohn singularity. For $\vect{r}$ and $\vect{v}$ parallel, this is:
\begin{multline}
\label{kohn_ahead}
\varphi_{\mathrm{kohn}}(\vect{r}) = i\frac{Q}{4\pi\epsilon_0}
\frac{1}{2r^2\lambda_{TF}^2}
\int_{-1}^1 \Biggl\{\frac{\sgn(\mu)}{k_{+}^3 [D(k_{+} v \mu, k_{+})]^2} \\
\times
\left[|\mu| -
\frac{2k_FM}{k_{+}}(1-\mu^2)\right]^2 \exp(ik_{+} r \mu)\Biggr\} d\mu,
\end{multline}
where $k_{+} = 2k_F(1 + M|\mu|)$. Only the $\emph{outer}$ surface of the
Kohn anomaly, i.e.\ $|\vect{k}| = k_{+},$\ 
contributes to this expression; the contribution of the inner surface vanishes because it
involves the integral $\int_{-\infty}^{\infty} e^{ix}/(x+i0) dx$, which
evaluates to zero by closing the contour in the upper half-plane. For $\vect{r}$
and $\vect{v}$ antiparallel, on the other hand, we get:
\begin{multline}
\label{kohn_behind}
\varphi_{\mathrm{kohn}}(\vect{r}) =
-i\frac{Q}{4\pi\epsilon_0} \frac{1}{2r^2\lambda_{TF}^2} \int_{-1}^1
\Biggl\{
\frac{\sgn(\mu)}{k_{-}^3 [ D(k_{-}v\mu, k_{-}) ]^2} \\
\times
\left[|\mu| + \frac{2k_FM}{k_{-}}(1-\mu^2)\right]^2 \exp(-ik_{-}r\mu) \Biggr\}
d\mu,
\end{multline}
where $k_{-} = 2k_F(1 - M|\mu|)$. This time only the $\emph{inner}$ surface,
i.e.\ $|\vect{k}| = k_{-}$, contributes.

The asymptotic formulas given in the body of the paper, 
Eqs.\
(\ref{asymptotic_quantum_ahead}--\ref{asymptotic_quantum_behind_stronger}),
follow from
Eqs.\ (\ref{kohn_ahead}) and (\ref{kohn_behind}) by changing the variable of
integration to $\xi = \mu k_{+}(\mu)$ [in \eqnref{kohn_ahead}], or
$\xi = \mu k_{-}(\mu)$ [in \eqnref{kohn_behind}], and considering the asymptotic
contribution as $r \to \infty$ from $\mu = \pm 1$ (the endpoints of the
integration), and from $\mu = 0$ (where the integrand is discontinuous).
In addition, the stronger $\sim 1/r^{2.5}$ Friedel oscillations arise from
points where
$d\xi/d\mu = 0$; this does not occur for
$\vect{r}$ and $\vect{v}$ parallel, whereas for $\vect{r}$ and $\vect{v}$
antiparallel and $M \geq 1/2$, it occurs at $\mu = \pm 1/(2M)$.
 
%
\end{document}